\documentclass[9pt,twocolumn,twoside]{osajnl}
\usepackage{siunitx}
\journal{ol} % Choose journal (ao, aop, josaa, josab, ol)
\setboolean{shortarticle}{true} 
\title{\SI[detect-weight]{0.24}{\tera\watt} Ultrabroadband, CEP-stable Multipass Ti:Sa Amplifier}

\author[1,2,*]{Mikayel Musheghyan}
\author[3]{Fabian L{\"u}cking}
\author[4]{Zhao Cheng}
\author[1]{Harald Frei}
\author[1]{Andreas Assion}

\affil[1]{High Q Laser GmbH, 1100 Vienna, Austria}
\affil[2]{University of Kassel, Institute of Physics, D-34132 Kassel, Germany}
\affil[3]{XARION Laser Acoustics GmbH, 1030 Vienna, Austria}
\affil[4]{Laboratoire d'Optique Apliquée, 91120  Palaiseau, France}
\affil[*]{Corresponding author: mikayel.musheghyan@spectra-physics.at}

\ociscodes{(140.3280) Laser amplifiers; (140.3590) Lasers, titanium; (260.7120) Ultrafast phenomena.}

\doi{\url{https://doi.org/10.1364/OL.44.001464}}

\begin{abstract}
We demonstrate a single-stage, multipass Ti:sapphire amplifier capable of delivering sub-\SI[detect-weight]{13}{\fs}, \SI[detect-weight]{3.2}{\milli\joule} pulses at \SI[detect-weight]{1}{\kHz} repetition rate. Gaussian filters are used to suppress the gain-narrowing effect, thereby enabling the achievement of an ultrabroadband flat-top spectrum with a FWHM~>~\SI[detect-weight]{130}{\nm}. The carrier-envelope phase (CEP) of the output pulses is actively stabilized via the feed-forward scheme (for the oscillator) and a fast \SI[detect-weight]{1}{\kHz} f-to-2f interferometer, which corrects the amplifier-induced CEP drift. The single-shot \SI[detect-weight]{75}{\hour} CEP measurement yielded rms noise of <\SI[detect-weight]{150}{\milli\radian}. 
\end{abstract}

\setboolean{displaycopyright}{true}

\begin{document}

\maketitle
\begin{figure}[htbp]
\centering
\fbox{\includegraphics[width=\linewidth]{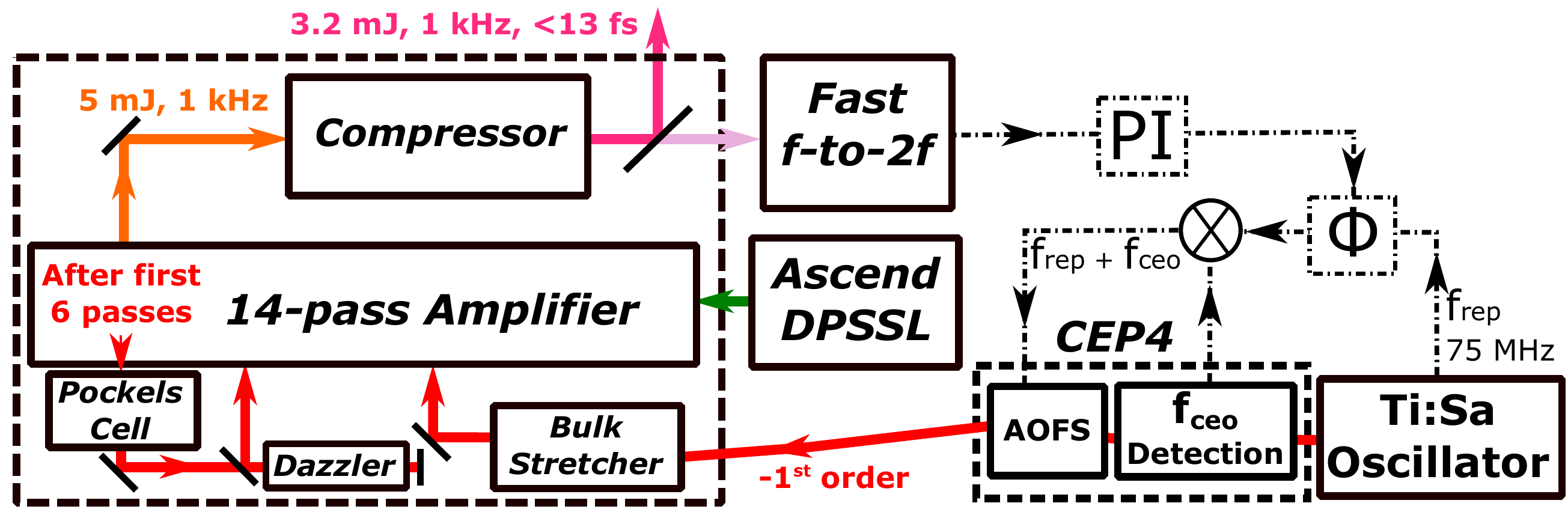}} 
\caption{Schematic of the experimental setup.\\ The CEP of the seed is stabilized via the CEP4 module. It is subsequently directed into the amplifier, which is pumped by the Ascend DPSSL. Output from the amplification stage (\SI[detect-weight]{5}{\milli\joule}) is compressed into \SI[detect-weight]{3.2}{\milli\joule}, sub-\SI[detect-weight]{13}{\fs} pulses. The CEP drift of the amplified pulses is stabilized by modifying the phase of the acousto-optic frequency shifter (AOFS) driver RF wave. This is possible by a feedback from the fast f-to-2f measurement: a proportional-integral controller PI generates the phase correction that is applied to the RF phase shifter $\Phi$.}
\label{fig:setup}
\end{figure}

%\section{Introduction}
After three decades of research and development, titanium-sapphire (Ti:Sa) chirped-pulse amplifiers~\cite{Strickland1985, Moulton1986} routinely deliver sub-\SI[detect-weight]{30}{\fs}, multi-millijoule pulses at repetition rates of up to several kilohertz. Nonlinear pulse compression techniques (e.g., as described in~\cite{Nisoli1996}) can further shorten the pulse durations and enable the generation of few-cycle pulses at a millijoule level. As a result, Ti:Sa amplifiers are indispensable tools in different fields of ultrafast physics, such as pump-probe spectroscopy, (soft) x-ray generation and plasma physics. Those amplifiers have also proven to be invaluable tools in the fields of higher-harmonic generation (HHG) and attosecond physics~\cite{Chang1997, Paul2001, Goulielmakis2010, Popmintchev2010, Baltuska2003, Lopez-Martens2005, Sansone2006, Krause1992, Krausz2009}. All of these applications (and many more) would benefit from shorter (<\SI[detect-weight]{25}{\fs}) pulses directly emitted from the amplifiers. In addition, certain experiments in the field of attosecond physics require the carrier-envelope phase (CEP) of these pulses to be stabilized. Hence, the combination of the CEP stability with shorter pulse durations from an amplifier may result in more robust, stable and simplified experimental setups. Apart from this, it should be mentioned as a side note that given the sufficient bandwidth, such ultrabroadband systems (with stabilized CEP) could potentially be used for direct pumping/seeding of optical parametric (chirped pulse) amplifiers (OP(CP)As)~\cite{Herrmann2009, Wilhelm1997, Cerullo1998, Baltuska2002}, as well as intra-pulse difference-frequency generation (DFG)~\cite{Vozzi2007}. \\
Although the bandwidth of the Ti:Sa crystal gain curve is sufficiently broad to enable the operation of a sub-\SI[detect-weight]{10}{\fs} oscillator, the gain-narrowing effect greatly complicates the generation of sub-\SI[detect-weight]{25}{\fs} millijoule-level pulses directly from an amplifier. The problem becomes more complex, if the CEP stabilization is required. These challenges are often easier to address with the OP(CP)As~\cite{Cerullo2003}, particularly taking into account the recent advancements in this technology~\cite{Andriukaitis2011, Budriunas2017}. Due to their (often) passive CEP stability~\cite{Baltuska2002PhysRev}, the broader gain bandwidth of the second-order nonlinear processes and the absence of a thermal load on the gain medium, OP(CP)As have proven to be a promising alternative to the Ti:Sa-based sources. However, there are certain obstacles to the practicality of OP(CP)A systems in terms of the simplicity, stability and scalability. Therefore, Ti:Sa-based broadband amplifiers may still be advantageous in a plethora of applications.\\
Over the past two decades, there have been various schemes to achieve shorter pulse durations for both regenerative and multipass amplifiers~\cite{Seres2003, AmaniEilanlou2008, Kalashnikov2016, Golinelli2017}. Most of the cited ultrabroadband systems, with the exception of~\cite{Golinelli2017}, do not feature CEP stabilization of the output pulses. This may be primarily attributed to the presence of a regenerative amplification stage. The problems in stabilizing the CEP of regenerative amplifiers arise due to the considerable CEP noise induced by the grating stretchers and compressors. The stretchers are necessary to achieve high stretching factors, which ensure that the B-integral values remain in the acceptable range. This is in contrast to the multipass amplifiers, which operate at smaller stretching factors. In multipass schemes, the seed can be stretched in a bulk material, thus avoiding (practically) any CEP noise from the stretcher. In addition, if the amplifier with a bulk stretcher supports an ultrabroadband spectrum, the stretching factor increases, making the entire system more reliable and less susceptible to damage. As a numerical example, \SI[detect-weight]{60}{\cm} of SF57 glass with group velocity dispersion (GVD) of \SI[detect-weight]{223.58}{\fs\squared/\mm}) will stretch Gaussian pulses with a FWHM~= \SI[detect-weight]{38}{\nm} from \SI[detect-weight]{25}{\fs} to \SI[detect-weight]{14.9}{\ps}, while pulses with a FWHM~= \SI[detect-weight]{72}{\nm} will be stretched from \SI[detect-weight]{13}{\fs} to \SI[detect-weight]{28.6}{\ps}.\\
Since most of the current ultrabroadband Ti:Sa systems feature several amplification stages and lack CEP stability, our main aim was to develop a compact CEP-stable multi-millijoule multipass amplifier with sufficient bandwidth to support sub-\SI[detect-weight]{15}{\fs} pulses. The multipass scheme was selected primarily for two reasons. First, as described above, the CEP stabilization is easier to implement when bulk stretchers are used. Second, the multipass scheme allows for the gain-narrowing compensation to be tailored to the specific gain regimes (small, intermediate and saturated). This enables greater versatility in comparison to the regenerative scheme, since the high-gain spectral components can be suppressed to a varying degree for each individual pass. We also aimed to investigate the pulse duration across the spatial profile of the output amplifier beam. Such spatially resolved measurements, which have not been reported in the researched literature, can be useful for verifying the homogeneity of the compression of the ultrabroadband spectrum.  \\
As a result of our efforts, we report in this letter a highly efficient, single-stage Ti:Sa multipass amplifier with sub-\SI[detect-weight]{13}{\fs}, \SI[detect-weight]{3.2}{\milli\joule} CEP-stable output pulses at \SI[detect-weight]{1}{\kHz} repetition rate. A single-shot CEP-measurement at the full repetition rate of the system was performed for over \SI[detect-weight]{75}{\hour} and resulted in a CEP rms noise of <\SI[detect-weight]{150}{\milli\radian}, making this system, to the best of our knowledge, the first amplifier with such an outstanding combination of short pulse duration, output energy and long-term CEP stability.\\
%\section{Experimental Setup}
%\label{sec:ExpSet}
Our experimental setup is depicted in Fig.~\ref{fig:setup}. The developed \SI[detect-weight]{1}{\kHz} multipass Ti:Sa amplifier consists of the following components: a few-cycle (<\SI[detect-weight]{7}{\fs}), \SI[detect-weight]{75}{\MHz}  seed oscillator (Element 2 prototype); an Ascend DPSSL pump; a CEP4 feed-forward CEP-stabilization module~\cite{Lucking2012} (f-to-2f variant) and a fast f-to-2f interferometer (all mentioned components are courtesy of Spectra-Physics). The latter is used for the amplifier CEP drift correction.
The oscillator output first passes through the CEP4 module. In order to measure the carrier-envelope offset (CEO) frequency $f_{CEO}$, 10\% of the input beam is directed (using a beam-splitter) to a photonic-crystal fiber for spectral broadening (to an octave) and a subsequent compact f-to-2f setup, which measures the frequency $f_{CEO}$. Frequency $f_{CEO}$ is then mixed with the oscillator repetition rate $f_{rep}$. The resulting $f_{CEO} + f_{rep}\approx$~85 MHz is used to drive an acousto-optic frequency shifter (AOFS), which diffracts the incoming oscillator beam (90\% of the original beam after the beamsplitter), thereby resulting in a CEP-stabilized $-1^{st}$ diffraction order. The AOFS efficiency at this driving frequency is 80\%. A more detailed description of this module is available in~\cite{Lucking2012}.\\
After the CEP4 module and before the amplification stage, the oscillator pulses are stretched in \SI[detect-weight]{60}{\cm} of SF57 glass to around \SI[detect-weight]{63}{\ps}. The seed power at the entrance of the amplification stage is \SI[detect-weight]{210}{\milli\watt}. The stage itself consists of 14 passes arranged in a "butterfly"-type configuration with two retro-reflectors. A detailed schematic of the multipass amplification stage is presented in~\cite{Sartania1997} (8-pass configuration) and~\cite{Hentschel2000} (10-pass configuration): the additional 4 passes (compared to \cite{Hentschel2000}) in our scheme are achieved by adding two more pairs of reflections on the retro-reflectors and the concave mirrors. The stage is pumped with \SI[detect-weight]{36.2}{\watt}; the pump focal spot being \SI[detect-weight]{1}{\milli\meter}. Seed beam size in the crystal in first 12 passes is approximately \SI[detect-weight]{300}{\micro\meter}, while in the 13th and 14th passes, the beam size of in the Ti:Sa crystal position is changed by telescopes to be closer to the pump focal spot size of \SI[detect-weight]{1}{\milli\meter}.\\
In the first 6 passes, the \SI[detect-weight]{75}{\MHz} oscillator pulse train is amplified to the microjoule level. The Pockels Cell selects pulses from the pulse train after the 6th pass at \SI[detect-weight]{1}{\kHz} rate. The Dazzler positioned after the Pockels Cell (in a double-pass configuration) is used to compensate for higher-order dispersion and to fine-tune the spectral shape. Subsequently, after 8 more passes, the pulses are amplified to \SI[detect-weight]{5}{\milli\joule}. By measuring the spectrum of the fluorescence of the amplifier crystal, it was devised that in this design the saturation effects become noticeable from 5-6th passes. In the last 3 passes the pulse energy is correspondingly \SI[detect-weight]{1}{\milli\joule}, \SI[detect-weight]{2.8}{\milli\joule}, \SI[detect-weight]{5}{\milli\joule}.\\
The compressor consists of a \SI[detect-weight]{1280}{lines\per\mm} transmission grating pair (Jenoptik AG) and 4 chirped mirrors (UltraFast Innovations GmbH, one bounce per mirror, -\SI[detect-weight]{250}{\fs^2} each) afterwards. Transmission gratings offer less thermal load compared to metallic reflective ones; however, chirped mirrors become necessary for the suppression of self-phase modulation inside the gratings.\\ 
The output of the system is \SI[detect-weight]{3.2}{\milli\joule}, sub-\SI[detect-weight]{13}{\fs} pulses at \SI[detect-weight]{1}{\kHz} repetition rate. Small portion of the beam (0.7\% of the total power) is guided to the fast f-to-2f interferometer for the CEP measurement. This interferometer features a Fringeezz (Fastlite) spectrometer, which allows single-shot CEP measurements at rates up to \SI[detect-weight]{10}{\kHz}. The fast f-to-2f, alongside with a proportional-integral (PI) controller and a phase shifter (within the CEP4 module), is part of the feedback scheme to correct the amplifier-induced CEP drift. Based on the CEP measurement done by the f-to-2f, a phase correction is generated by the PI controller and applied to the phase shifter. This modifies the phase of the AOFS driver frequency, compensating for the amplifier-induced CEP drift. Due to the system's repetition rate and the Nyquist theorem, the drift correction is theoretically limited to frequencies below \SI[detect-weight]{500}{\Hz}. A more detailed description is available in~\cite{Lucking2014}.\\
In comparison to a standard millijoule-class multipass <\SI[detect-weight]{30}{\fs} Ti:Sa amplifier (FWHM~> \SI[detect-weight]{40}{\nm}), the following changes have been performed to achieve and maintain a reliable ultrabroadband (FWHM~> \SI[detect-weight]{130}{\nm}) operation:\\
1. Use of custom filters with a Gaussian-shaped reflectance profile. The profile is based on the shape of the gain curve of Ti:Sa. These filters are positioned near the retro-reflectors (similar to the position of the shaping filter in~\cite{Hentschel2000}) in 12 out of the total 14 passes (last two passes, in which the pulses have high energy are left unfiltered). This greatly suppresses the gain-narrowing effect. By slightly tilting the Gaussian filters (thus changing the incident beam angle), it is possible to tune the spectrum and guarantee the broadest bandwidth.\\
2. Use of the Dazzler in a double-pass configuration. Due to the large bandwidth, two passes are required to compensate for the higher-order dispersion coefficients. The double-pass scheme also minimizes the spatial chirp (hence, one mirror is used instead of a retro pair, as it is shown in Fig.~\ref{fig:setup}). This chirp occurs because different wavelengths are diffracted at different positions inside the Dazzler~\cite{Kaplan2002}. It should be noted that although spectral shaping in principle can be performed only by using the Gaussian filters, fine-tuning of the spectrum via the Dazzler guarantees more robust performance and easier alignment.\\ 
3. Use of silver mirrors to preserve the generated bandwidth. Most of the mirrors in the setup and inside the amplifier were replaced by silver ones. High reflective dielectric mirrors were left only in the positions where a higher damage threshold was crucial.\\
The efficiency of the system (direct output of the amplification stage before the compression over the total pump power) is 13.8\%, while the efficiency of a \SI[detect-weight]{30}{\fs} system is 20\%. This is mainly due to the losses induced by the Gaussian filters and the double-pass Dazzler. The compressor efficiency also decreased from 80\% to 64\%, since the grating transmission near the edges of the spectrum is lower.\\
%\section {Results}
%\subsection {Spectrum and Pulse Duration}
%\label{sec:SpecDur}
The compressed output pulses of the system exhibit a flat-top spectrum with a FWHM~= \SI[detect-weight]{134}{\nm} and \SI[detect-weight]{3.2}{\milli\joule} energy. The pulse duration is $\leq$\SI[detect-weight]{13}{\fs} as measured by both the Wizzler (Fastlite~\cite{Oksenhendler2010}) and the d-scan (Sphere Ultrafast Photonics~\cite{Miranda2012}). Before the pulse duration measurements, the higher-order dispersion coefficients were optimized using the Wizzler-Dazzler feedback loop.
\begin{figure}[htbp]
\centering
\fbox{
\includegraphics[width=\linewidth]{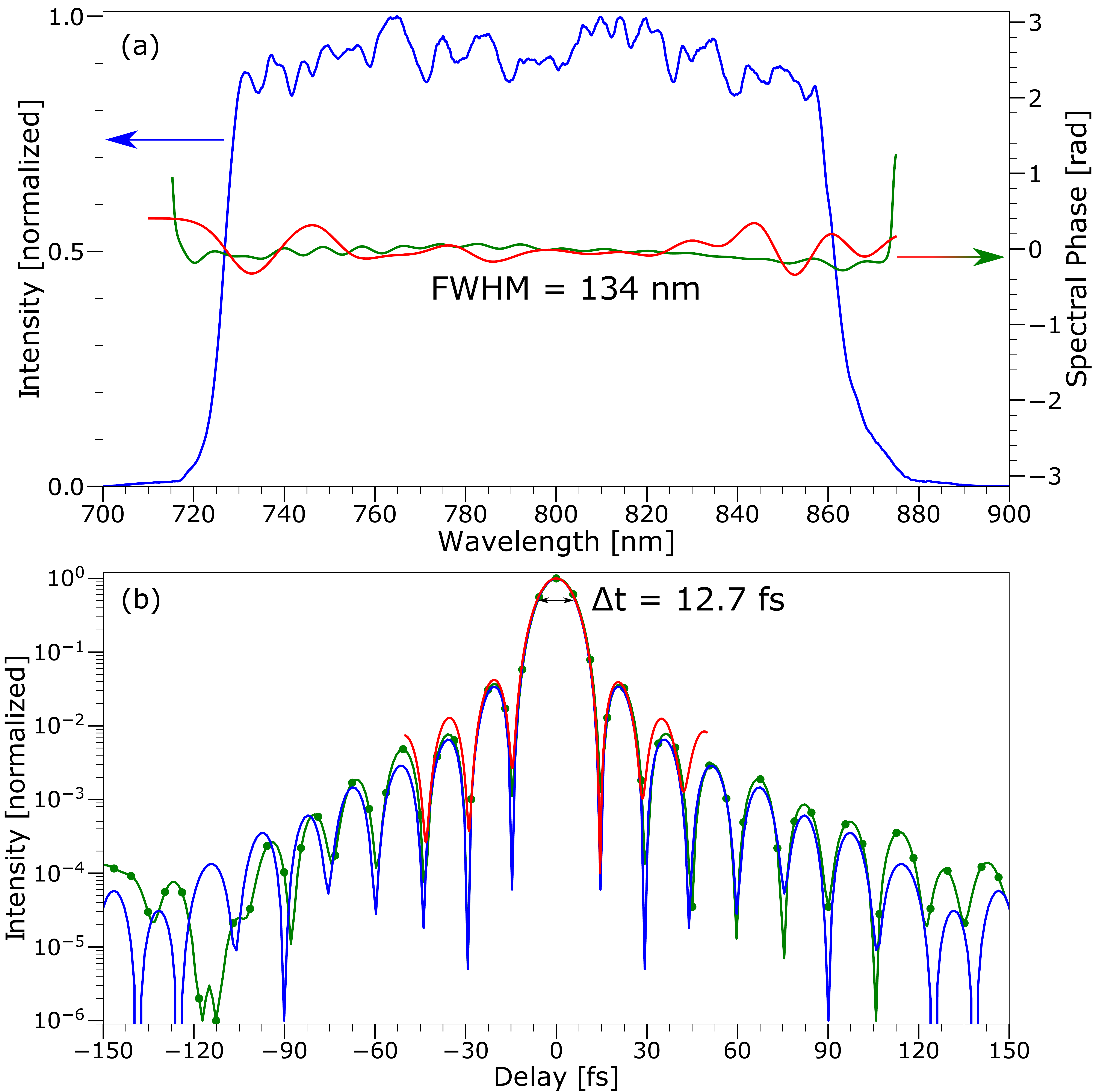}}
\caption{Pulse duration measurements.\\
\textbf{(a) Spectral domain.} \textit{Solid blue line} - measured output spectrum; \textit{solid green line} - spectral phase measured by the Wizzler; \textit{solid red line} - spectral phase measured by the d-scan.\\
\textbf{(b) Temporal domain.} \textit{Solid blue line} - FTL pulse; \textit{solid green line} - Wizzler measurement; \textit{solid red line} - d-scan measurement. Wizzler-Dazzler feedback allows for the compression nearly to the FTL.}
\label{fig:pulsechar}
\end{figure}
The spectrum, Wizzler and d-scan measurements are shown in Fig.~\ref{fig:pulsechar}. The spectral phase in the case of the Wizzler measurement exhibits practically no oscillations, while in the case of the d-scan result features slight oscillatory behavior. The underlying reasons for this are not clear; however, the effect on the pulse duration is negligible. Temporal pulse shapes retrieved both by the Wizzler and the d-scan are similar (Wizzler measurements were on average 1.6\% shorter) and close to the Fourier transform limited (FTL) shape.\\
\begin{figure}[htbp]
\centering
\fbox{\includegraphics[width=\linewidth]{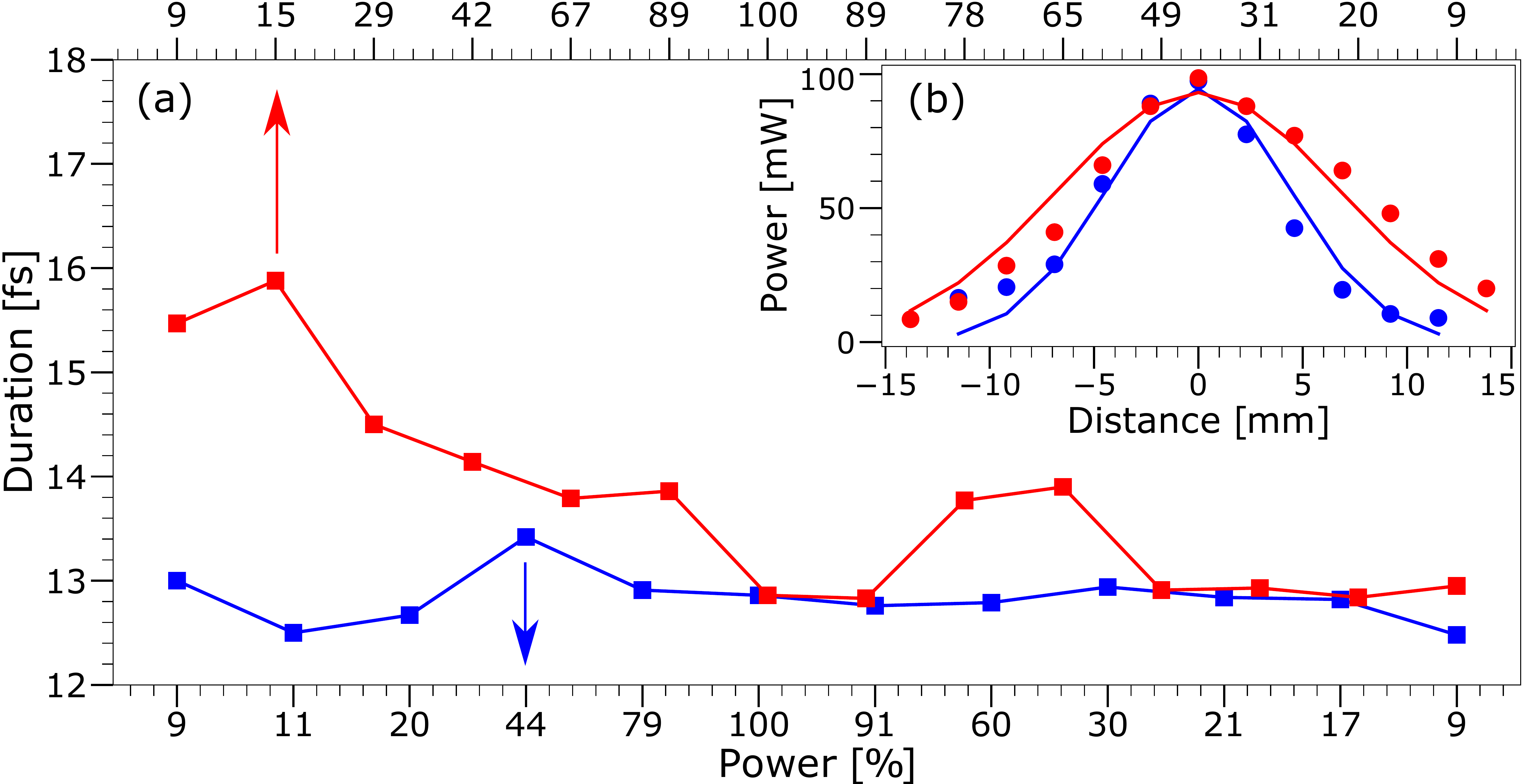}}
\caption{
Spatially resolved measurements.\\
\textit{Solid blue lines} - horizontal beam axis, along which spatial chirp from the Dazzler may occur; \textit{Solid red lines} - vertical beam axis, along which the compression gratings diffract.\\
\textbf{(a) Spatially resolved pulse duration.} The X-axes represent the percentage power relative to the beam center, i.e., 100\% power corresponds to the middle of the beam, and the power/percentage decreases towards the spatial edges of the beam. Our measurements show homogeneous duration along the horizontal beam axis. The vertical beam axis has a longer duration near one of the edges due to the limited size of the gratings.\\
\textbf{(b) Deduced spatial profile.} The Gaussian curve is fitted to the measured spatial profile of the beam. Points in (b) correspond to points in (a), e.g., the zero point in (b) the 100\% point in (a). The difference for horizontal and vertical beam axes can be attributed to the thermal lensing effect within the amplifier crystal.\\}
\label{fig:spatdep}
\end{figure}
The spatial dependence of the pulse duration was investigated for both beam axes and the results are plotted in Fig.~\ref{fig:spatdep}. The beam axes are labeled based on the possible spatial chirp sources: the horizontal axis (blue lines) is the axis in which the Dazzler can induce a spatial chirp, while the vertical axis (red lines) is the diffraction axis of the compressor gratings. A \SI[detect-weight]{2.3}{\mm} aperture fixed to a micrometer stage was moved across the spatial profile of the collimated beam along both the horizontal and vertical axes. The collimated beam size ($1/e^2$) was \SI[detect-weight]{17.6}{\mm} in horizontal and \SI[detect-weight]{27.1}{\mm} in vertical axis (calculated from the fittings in Fig.~\ref{fig:spatdep}~(b)). The pulse duration measurements were recorded using the d-scan. The X-axes in Fig.~\ref{fig:spatdep}~(a) correspond to the power of the beam after the \SI[detect-weight]{2.3}{\mm} aperture, expressed in percentage, where 100\% is the power after the aperture in the center of the beam. The Y-axis corresponds to the measured pulse duration.\\
The horizontal beam axis measurement shows that the spatial chirp induced by the Dazzler is minimal and the pulse duration is homogeneous across the entire profile. The vertical beam axis measurement shows an increased duration near one of the edges of the beam. The reason is the limited geometrical dimension of the gratings, which cut out a part of the spectrum near one of the (spatial) edges of the beam. This may be remedied by enlarging the size of the gratings. The effect is not extreme and after 20\% point the duration is again sub-\SI[detect-weight]{15}{\fs}. The slight inhomogeneity near the central part of the curve remains unresolved.\\
%\subsection {Amplifier CEP-Stability}
%\label{sec:CEP}
The CEP measurements were performed for \SI[detect-weight]{75}{\hour}. The CEP of the amplified pulses was recorded single-shot, at a \SI[detect-weight]{1}{\kHz} acquisition rate. The power spectral density (PSD) curve of this dataset was evaluated by dividing the total data into overlapping segments, calculating a modified periodogram for each segment, and averaging them afterwards (Welch's method~\cite{Welch1967}). The integrated phase noise (IPN) was derived from the resulting PSD curve.
\begin{figure}[htbp]
\centering
\fbox{\includegraphics[width=\linewidth]{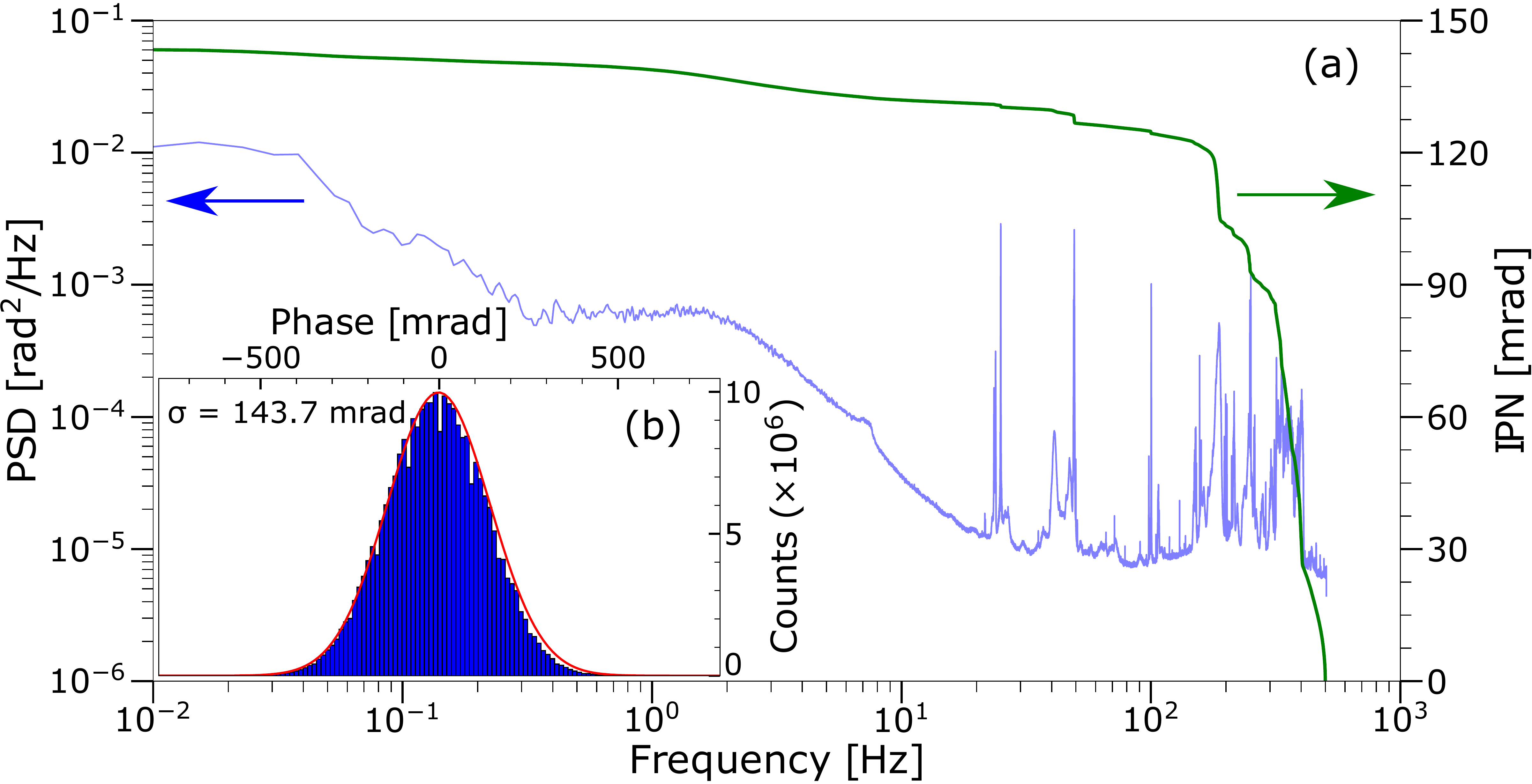}}
\caption{Results of \SI[detect-weight]{75}{\hour} CEP measurement.\\
\textbf{(a) PSD and IPN curves.} \textit{Solid blue line} - power spectral density; \textit{solid green line} - integrated phase noise. The main contributions to the CEP noise occur at frequencies above \SI[detect-weight]{200}{\Hz}. The resolution of the measurement is 6-\SI[detect-weight]{10}{\milli\radian} from the product description of the Fringeezz~\cite{Fringeezz2018}.\\
\textbf{(b) CEP histogram.} A Gaussian (\textit{solid red line}) is fitted with the $\sigma$ parameter displayed on the top left.}
\label{fig:cep}
\end{figure}
The results are depicted in Fig.~\ref{fig:cep}. After \SI[detect-weight]{75}{\hour}, the CEP single-shot noise is exceptionally low at \SI[detect-weight]{144}{\milli\radian}. This indicates that the modifications performed in order to achieve shorter pulse durations do not compromise the CEP stability of the system. \\
%\section{Conclusions}
In this study, we demonstrated an ultrabroadband multi-millijoule Ti:Sa multipass amplifier with unprecedented long-term CEP stability. By performing the spectral shaping in different passes using the Gaussian filters, we were able to achieve significant suppression of the gain-narrowing effect and attain a flat-top spectrum with a FWHM~> \SI[detect-weight]{130}{\nm}. The Wizzler-Dazzler feedback loop enabled to compress the pulses nearly to the FTL, resulting in sub-\SI[detect-weight]{13}{\fs} pulses. The pulse duration was measured across the spatial beam profile to inspect the homogeneity of the compression.\\
Based on our investigations, the following steps may further enhance the output parameters. First, the output energy can be scaled up by at least a factor of 2-3 using a booster stage. Low amplification factor allows for the gain-narrowing effect compensation using only the Dazzler. Second, larger transmission gratings, featuring high transmission >\SI[detect-weight]{150}{\nm} (available from the company Gitterwerk GmbH) can further increase the compressed output energy. Lastly, it may be possible to achieve shorter pulse duration using a Dazzler with a broader diffraction bandwidth. \\
In terms of the output parameters, the developed ultrabroadband single-stage Ti:Sa amplifier can be considered a promising step in filling the gap between more complex (and often more expensive) OP(CP)A systems and the standard Ti:Sa sources.\\
%\section{Funding Information}
{\fontfamily{phv}\selectfont{\fontsize{10pt}{12pt}\textbf{Funding}}}. H2020-MSCA-ITN-2014-641789-MEDEA (Marie Skłodowska-Curie Innovative Training Networks). 
% Bibliography
\bibliography{Bibliography}
\bibliographyfullrefs{Bibliography}
% Full bibliography added automatically for Optics Letters submissions; the following line will simply be ignored if submitting to other journals.
% Note that this extra page will not count against page length

\end{document}